\documentclass{aa}  

\usepackage{graphicx}
\usepackage{txfonts}
\usepackage{natbib}
\usepackage{cases}
\usepackage{comment}
\usepackage{soul}

\usepackage{xcolor}
\usepackage{subcaption}
\usepackage[normalem]{ulem}
\usepackage[colorlinks=true,allcolors=blue]{hyperref}
\usepackage{graphicx}
\usepackage{txfonts}

\begin{document}

   \title{Collisional excitation of propyne (CH$_3$CCH) by He atoms
}


   \author{M. Ben Khalifa
           \fnmsep\thanks, B. Darna 
          \and
          J. Loreau
          }

   \institute{KU Leuven, Department of Chemistry, Celestijnenlaan 200F, B-3001 Leuven, Belgium.\\
              \email{malek.benkhalifa@kuleuven.be, jerome.loreau@kuleuven.be}
                      }


 
  \abstract
   {A detailed interpretation of the detected emission lines of environments in which propyne (or methyl acetylene, CH$_3$CCH) is observed requires a knowledge of its collisional rate coefficients with the most abundant species in the interstellar medium, He or H$_2$.}
   {We present the first three-dimensional potential energy surface (3D-PES) for the CH$_3$CCH-He molecular complex, study the dynamics of the collision, and report the first set of rate coefficients for temperatures up to 100 K for the collisional excitation of the lowest 60 \textit{ortho} rotational levels and 60 \textit{para} rotational levels of CH$_3$CCH by He atoms.}
   {We computed the 3D-PES with the explicitly correlated coupled-cluster with single-, double-, and perturbative triple-excitation method,  in conjunction with the augmented correlation-consistent triple zeta basis set (CCSD(T)-F12a/aug-cc-pVTZ). The 3D-PES was fitted to an analytical function. Scattering computations of pure rotational (de-)excitation of CH$_3$CCH by
collision with He atoms were performed and the state-to-state cross sections were computed using the close coupling method for
total energies up to 100 cm$^{-1}$ and with the coupled states approximation at higher energy for both \textit{ortho} and \textit{para} symmetries of CH$_3$CCH.}
   {The PES obtained is caracterized by a large anisotropy and a potential well depth of 51.04 cm$^{-1}$. By thermally averaging the collisional cross sections, we determined quenching rate coefficients for kinetic temperatures up to 100 K. A strong even $\Delta j$ propensity rule at almost all collision energies exists for CH$_3$CCH-He complex. To evaluate the impact of rate coefficients in the analysis of observations, we carried out non-LTE radiative transfer computations of the excitation temperatures and we demonstrate that LTE conditions are typically not fulfilled for the propyne molecule.}
   {}


   \maketitle
%
\section{Introduction}
Over the past decades, methyl acetylene (or propyne, CH$_3$CCH) has received significant attention from the astrophysical and astrochemistry communities due to its important role in astrochemical processes \citep{mebel2017formation,herbst2017synthesis}
such as its role as a precursor in the formation of several polycyclic aromatic hydrocarbons (PAHs)\citep{parker2017formation}.\\
The earliest tentative detection of propyne in the  interstellar medium (ISM) was reported toward the Sgr B2 molecular cloud by \cite{buhl1973detection}. Later, 
propyne was observed by \cite{lovas1976millimeter} towards Orion A and Sagittarius B2 through the 5$_0 \rightarrow 4_0$ rotational line at 85 GHz.
In the following years, CH$_3$CCH was identified in various astronomical environments:
low mass star-forming regions \citep{van1995molecular}, photodissociation regions through the Horsehead nebula with fractional abundances of 10$^{-9}$ with respect to molecular hydrogen \citep{gratier2013iram,guzman2014chemical,hickson2016methylacetylene}, massive young stellar objects \citep{fayolle2015complex}, circumstellar envelopes of evolved stars 
\citep{agundez2008detection} and even toward extragalactic sources such as M82, NGC 253, and NGC 1068 
\citep{mauersberger1991dense,qiu2020lambda}. It has also been detected in cold and dense cores \citep{vastel2014origin,gratier2016new} and even toward a planetary nebula, as recently reported by \cite{schmidt2019exotic}.
As a matter of fact, the widespread detection of propyne even extends to planetary atmospheres :  
in our solar system, CH$_3$CCH has also been detected on Jupiter, Saturn, Uranus as well as in the atmosphere of Titan 
\citep{fouchet2000jupiter,de1997first,burgdorf2006detection,teanby2009titan}.\\
The formation pathways of propyne in the gas phase have been widely studied and it has been demonstrated that there are no efficient synthetic pathways in gas-phase available to reproduce the abundances of CH$_3$CCH in the cold molecular clouds \citep{hickson2016methylacetylene}.
It was proposed that the principal source of propyne in the gas phase is via ion-molecule reactions with C$_2$H$_2^+$ as a precursor \citep{schiff1979ion},  neutral-neutral reactions such as CCH + CH$_4 \rightarrow$ CH$_3$CCH + H \citep{turner1999physics} 
as well as dissociative recombination reactions involving larger hydrocarbons \citep{calcutt2019alma}.\\
Those pathways were not capable to reproduce the observed abundance of CH$_3$CCH in the astrophysical media. For this reason, surface reactions occurring on interstellar grains were additionally studied \citep{hickson2016methylacetylene,guzman2018chemistry}.
Propyne molecules are believed to adhere to the cold dust grains and then undergo hydrogenation at cold molecular cloud temperature of about 10 K. 
CH$_3$CCH was demonstrated to form through the hydrogenation of C$_3$ radical which could be further hydrogenated to propene.
Nevertheless, the model including grain surface and gas phase reactions was not able to reproduce the observed methyl acetylene abundance by more than one order of magnitude.
This failure to reproduce the abundance of propyne demonstrate that more crucial formation pathways are still unknown at low temperatures \citep{hickson2016methylacetylene,guzman2014chemical,oberg2013spatial}.\\
CH$_3$CCH is a symmetric top molecule and among the most efficient thermometers available. For a given $j$, the rotational levels associated to propyne are split into several levels with different $k$ quantum numbers. Consequently, there are multiple rotational transitions with different $k$ quantum numbers that cover a wide energy range above the ground state, but that are closely spaced in frequency and can therefore all be observed in the same bandwidth: that is, many transitions can be observed simultaneously using the same antenna settings and sensitivities thereby reducing the calibration uncertainties. The dipole moment of propyne is parallel to the C$_3$ axis, hence, radiative transitions with $\Delta k \neq$ 0 are forbidden: different $k-$components are connected only via collisional process, which makes their relative population sensitive to the kinetic temperature.\\
The interstellar medium where propyne generally resides is characterized by a very low density, and therefore the populations  of its rotational states are usually not in  local thermodynamic equilibrium (LTE). Therefore, the analysis of intensities of spectroscopic lines of CH$_3$CCH requires the application of a radiative transfer model. Such a model requires spectroscopic data for radiative transfer rates, which are usually available, as well as collisional rate coefficients for rotational transitions induced by collisions with the dominant interstellar species, usually He or H$_2$, which are aften not available. Consequently, these collisional rates must be calculated in quantum scattering computations using a precise potential  energy surface (PES) for the interaction  of the molecule with He or H$_2$.\\
To the best of our knowledge, no collisional state-to-state rate coefficients are available in the literature and the present set of collisional rate coefficients is the first one obtained by fully quantum methods.\\
In the present paper, we study the collisional excitation
of the rigid symmetric top molecule CH$_3$CCH with He
atoms for temperature up to 100 K.
The overall structure of this paper is as follows: section~\ref{section:2} describes the \textit{ab-initio} computation of the 3-dimensional CH$_3$CCH-He intermolecular potential, the analytical representation of the interaction potential is defined and the potential is illustrated. In Sect.~\ref{section:3} we describe the study of the dynamics, where we
illustrate inelastic cross sections in CH$_3$CCH-He collisions.
We discuss rate coefficients for CH$_3$CCH-He collisions in Sect.~\ref{section:4}. We
analyze the impact of our rate coefficients by performing a radiative
transfer calculation for typical interstellar conditions in Sect.~\ref{section:5}.
Conclusions and future outlooks are drawn in Sect.~\ref{section:6}.

\section{Potential energy surface}\label{section:2}
\begin{figure}
\centering
	\includegraphics[width=0.6\columnwidth]{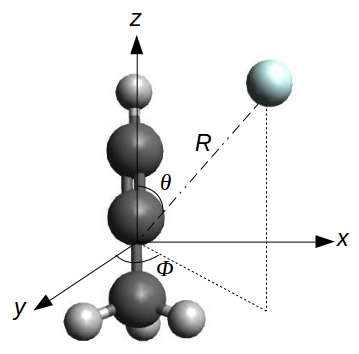}
	\caption{Jacobi coordinates defining the geometry of the CH$_3$CCH-He van der Waals complex. The origin of the 
 reference frame is at the CH$_3$CCH center-of-mass. The angles 
 are defined so that for $\theta$= 0$^{\circ}$, He approaches 
 along the C-H end of the molecule and for $\phi$= 60$^{\circ}$,
 the He atom lies in the plane defined by CHCH atoms on the side 
 of the H atom.} \label{Jacobi}
\end{figure}

In this section, we aim to compute the first accurate potential energy surface between a symmetric top molecule, CH$_3$CCH ($X^1A_1$), with a structureless helium atom ($^1$S) in their respective electronic ground states.\\
The 3D-PES was computed as a function of three Jacobi coordinates ($R$,$\theta$,$\phi$). The origin of the coordinate system is the centre of mass of CH$_3$CCH molecule and the $Z$ axis lies along its C$_3$ symmetry axis. $R$ denotes the intermolecular separation between the center of mass of CH$_3$CCH and He atom and $\theta$ and $\phi$ define the orientation of the He atom. The coordinates defining the geometry of the CH$_3$CCH-He system are presented in Fig.~\ref{Jacobi}. 
For $\phi$=0$^{\circ}$, the He atom is in the plane formed by the axis of molecule and one of the C–H bonds, while for $\phi$=60$^{\circ}$, the He atom is situated at equal distance between two hydrogen atoms.\\
The propyne molecule was assumed to be rigid in the computation of the interaction potential of CH$_3$CCH–He where the internal coordinates of CH$_3$CCH were frozen to the average geometry of the $\nu$=0 vibrational state as follows:
$r$(H--C)=1.061~\AA, $r$(C$\equiv$C)=1.215~\AA, $r$(C--C)=1.458~\AA, $r$(C--H)=1.089~\AA~and $\angle$(HCC)= 110.7$^{\circ}$ \citep{el2001vibrational}. \\
Explicitly correlated coupled cluster computations with inclusion of single, double, and perturbatively triple excitations [CCSD(T)-F12a] \citep{knizia2009simplified} in conjunction with the augmented correlation consistent polarized valence triple zeta (aug-cc-pVTZ) basis sets of Dunning \citep{dunning1989gaussian} were used to calculate $ab$-$initio$ points on the CH$_3$CCH-He interaction potential and the {\small MOLPRO} code \citep{werner2015molpro} was used to perform these computations in the C$_1$ symmetry group.\\
The performance of such an approach was already tested for CH$_3$CN-He, CH$_3$NC-He \citep{ben2022interaction} as well as SiH$_3$CN-He \citep{naouai2021inelastic}, and it was proven that CCSD(T)-F12 method supplemented by the aVTZ  basis set give an excellent accuracy for interactions involving rare gases.\\
A counterpoise correction of \citet{boys1970calculation} was considered to correct for the basis set superposition error (BSSE) according to the following expression:
\begin{equation}\label{eq:1}
 V(R,\theta,\phi)=V_{\rm{Mol-He}}(R,\theta,\phi)-V_{\rm{Mol}}(R,\theta,\phi)-V_{\rm{He}}(R,\theta,\phi)
\end{equation}
where $V_{\rm{Mol-He}}$ is the global electronic energy of CH$_3$CCH-He system, and the last two terms are the energies of the two fragments, all performed using the full basis set of the total system.\\
The ranges of variation used to compute the PES was constructed as follows. For the angle $\theta$, the grid is made up of 19 values from 0$^{\circ}$ to 180$^{\circ}$, while for the angle $\phi$, we used 7 values from 0$^{\circ}$ to 60$^{\circ}$ uniformly distributed by steps of 10$^{\circ}$. All other geometries are obtained by symmetry.
Finally, for the intermolecular distance $R$, we used a grid of 60 points (4 $\leq R \leq$ 50 a$_0$). We used a spacing of 0.2 a$_0$ in the short-range, while in the long-range, the spacing is increased progressively. These parameters lead to a set of 7980 energy points.\\
We take into consideration the inconsistency in size of the CCSDT(T)-F12a method, thus, the PES computed with this approach was shifted by subtracting the asymptotic value of the PES at $R$ = 50 a$_0$, which is equal to $-$5.3 cm$^{-1}$.
\subsection{Analytical fit}
\begin{figure}
\centering
	\includegraphics[width=0.99\columnwidth]{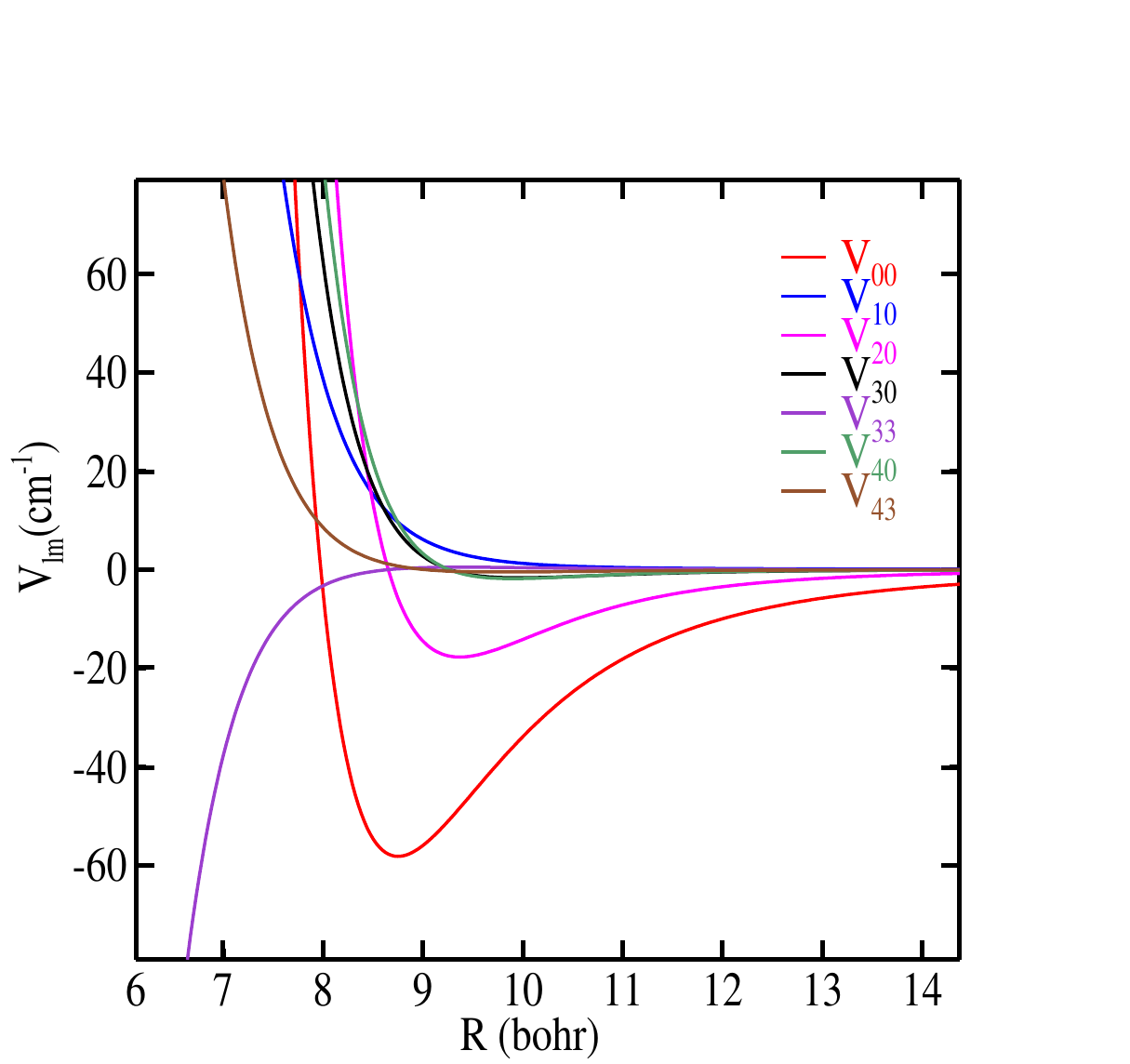}
	\caption{Dependence on $R$ of the first $V_{lm}(R)$ components for CH$_3$CCH-He with 0 $\leq l \leq$ 4.}\label{coeff}
\end{figure}

It is useful for the computation of the matrix elements of the interaction between scattering basis functions in time-independent quantum method to use the functional form presented below:
\begin{equation}
V(R,\theta,\phi)=\sum_{l=0}^{l_{max}}\sum_{m=0}^{l}V_{lm}(R)\frac{Y_{l}^{m}(\theta,\phi)+(-1)^{m}Y_{l}^{-m}(\theta,\phi)}{1+\delta_{m,0}}
\end{equation}
where $V_{lm}(R)$ and $Y_{l}^{m}(\theta,\phi)$ are the radial coefficients to be computed and the normalized spherical harmonics respectively, and $\delta_{m,0}$ is the Kronecker delta symbol.\\
Due to the C$_{3v}$ symmetry of the propyne molecule, the values of $m$ are restricted to terms for which $m$ is multiple of 3 ($m=3n$ , $n$ integer) in the expansion.\\
From a PES grid containing (19,7) values of ($\theta$,$\phi$),
we could include radial coefficients up to $l_{max}$=18 and $m$=18.
The total number of expansion coefficients is equal to 70 with a final accuracy better than 1 cm$^{-1}$ for $R \geq$ 5 bohr.
Finally, a least-square interpolation was used over the entire range of intermolecular distances to supply continuous expansion coefficients suitable for scattering computations. For $R$ $\geq$ 30 bohr, we extrapolated the long-range potential using an inverse exponent expansion implemented in the {\small MOLSCAT} \citep{molscat95} computer code.\\
We present in Fig.~\ref{coeff} the dependence on $R$ of the radial coefficients up to $l$= 4. 
Here, $V_{00}$ represents the isotropic potential, responsible for elastic collisions, while terms with $l \geq$ 1 describe the anisotropic part of the PES responsible for inelastic collisions.
As one can see, the largest (in magnitude) of the anisotropic terms
($l$ > 0) is associated to $l$ = 2, a term that is reponsible for transitions with $\Delta j$=2. 
This should have an effect on the propensity rules in the rotational excitation, as discussed in more detail below.

\subsection{Analysis of the PES}
\begin{figure*}
\centering
{\label{c}\includegraphics[width=.49\linewidth]{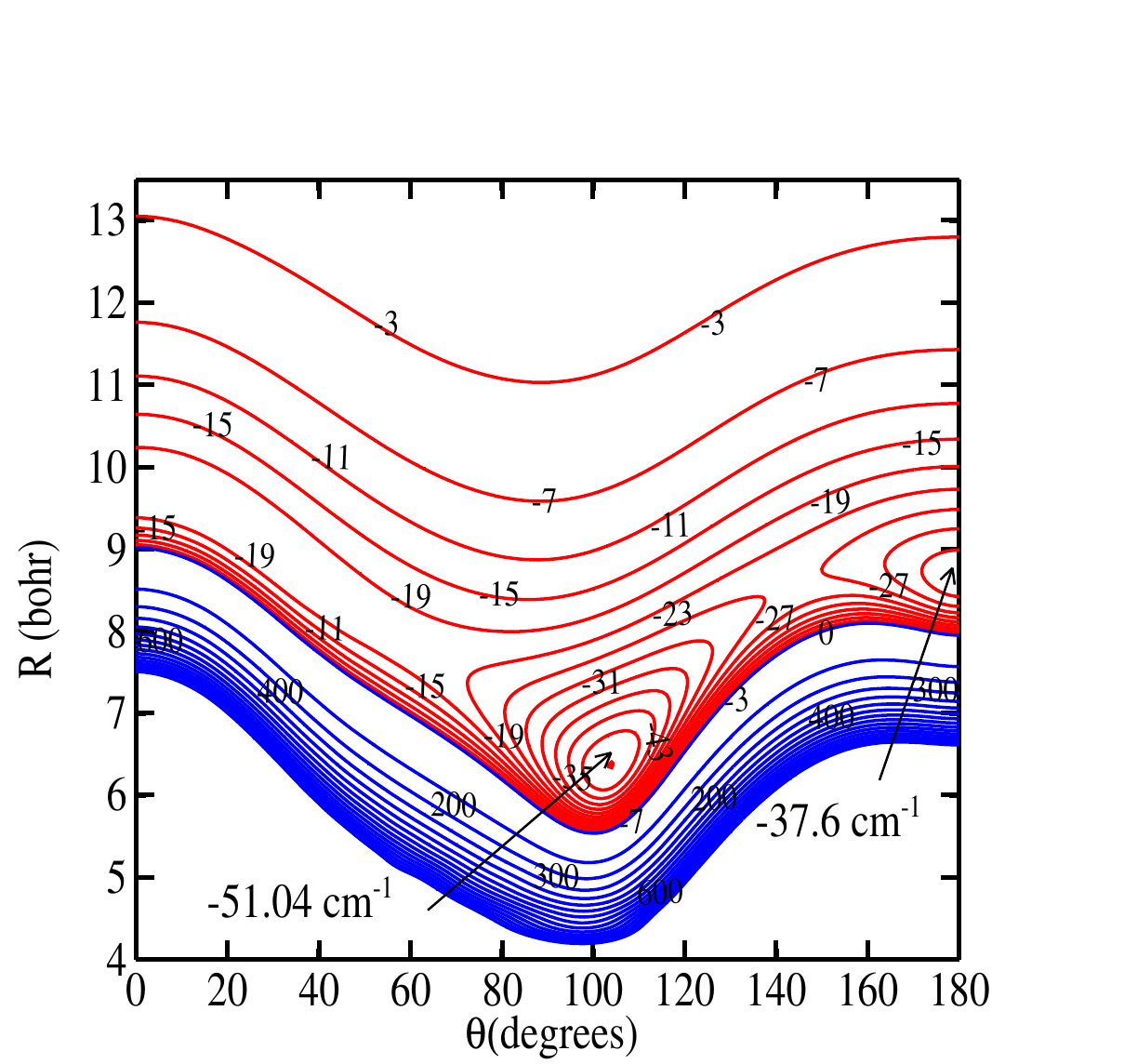}}
{\label{c}\includegraphics[width=.49\linewidth]{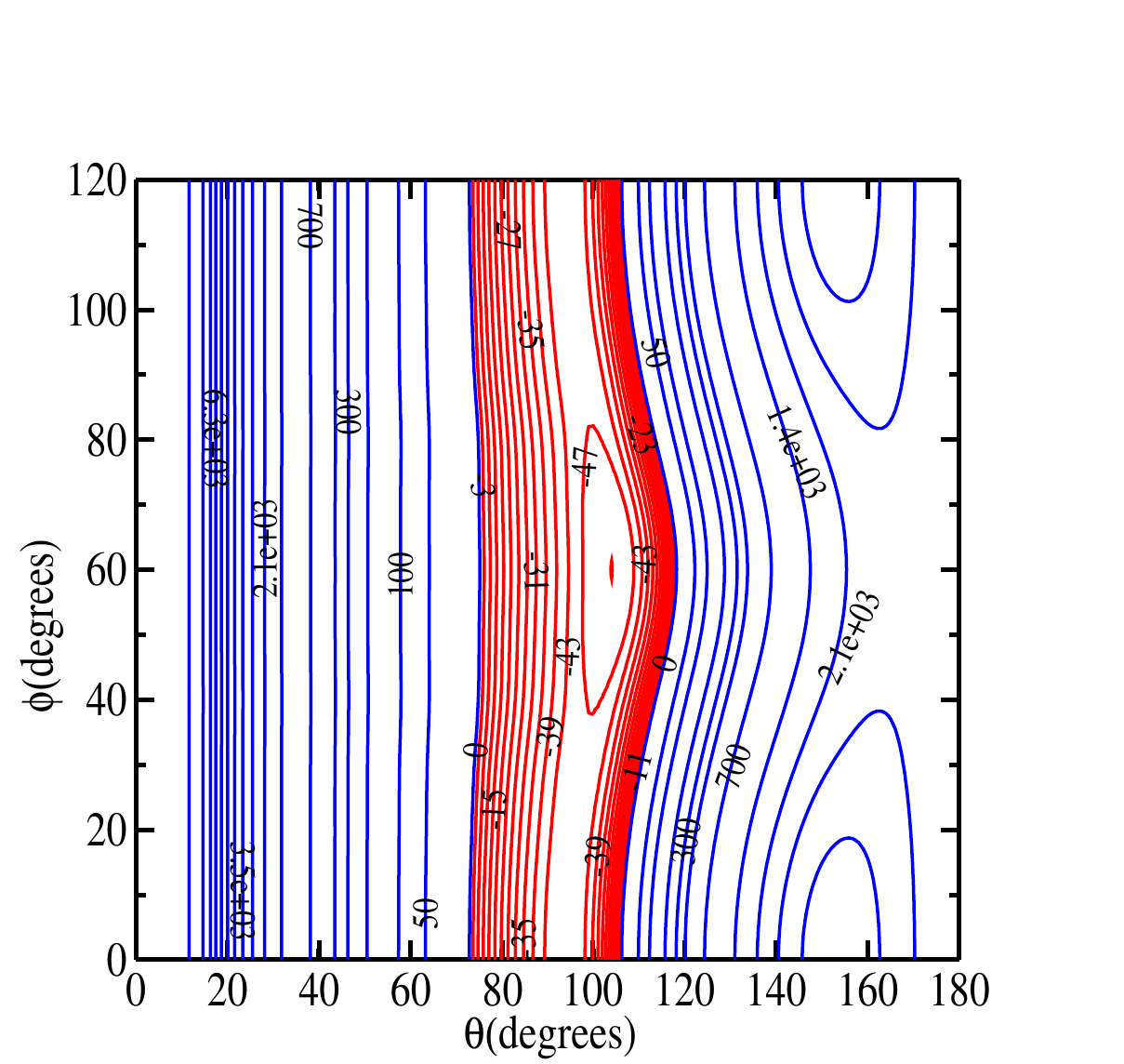}}
\caption{Two-dimensional contour plots of the interaction 
potential of the CH$_3$CCH-He van der Waals complex.The left 
panel depicts the 3D-PES as a function of $\theta$ and $R$ at 
$\phi$=60$^{\circ}$,  while the right panel shows the PES as a
function of $\phi$ and $\theta$ at $R$= 6.6a$_0$. For each 
panel, the blue (red) contours represent the positive (negative)
parts of the potential (in unit of cm$^{-1}$).}
\label{3D-PES} 
\end{figure*}

 The global minimum of the PES, which describe the main interaction of CH$_3$CCH-He, is located at the equilibrium intermolecular separation $R_e$=6.3 bohr with a well depth of $D_e$= 51.04 cm$^{-1}$. The configuration of the global minimum is defined by the angles $\phi=$ 60$^{\circ}$ and $\theta=$ 104$^{\circ}$, while a local minimum of 37.6 cm$^{-1}$ is located at $R=8.6$ bohr and $\theta=$ 180$^{\circ}$. The global and local minima are separated by a barrier of $-$27.07 cm$^{-1}$ that is located at $\theta=$ 137$^{\circ}$ and $R=8.35$ bohr.\\
The internal rotation along the $\phi$ coordinate is almost free:
the energy of the minimum increases between $\phi=0^{\circ}$ and $\phi=30^{\circ}$ (from $-46.8$ cm$^{-1}$ at $\theta=94^{\circ}$ to $-46.6$ cm$^{-1}$ at $\theta=98^{\circ}$) and then decreases between $\phi=30^{\circ}$ and $\phi=60^{\circ}$ (from $-46.6$ cm$^{-1}$ at $\theta=98^{\circ}$ to $-51.04$ cm$^{-1}$ at $\theta=104^{\circ}$). This behavior was already observed for other systems with C$_{3v}$ symmetry such as CH$_3$CN-He and CH$_3$NC-He \citep{ben2022interaction}, and is due to the elongated shape of the propyne molecule.\\ 
 We note that the PES for CH$_3$CCH-He share the same qualitative behavior previously observed for molecules with a threefold symmetry axis interacting with rare gas atoms  \citep{Gubbels2012, Loreau2014b, loreau2015scattering, ben2022interaction, naouai2021inelastic} where the global minimum occurs at $\phi$=60$^{\circ}$, that is, with the helium atom located between two hydrogen atoms.
Because of the C$_{3v}$ symmetry of the CH$_3$CCH molecule, there are two more symmetry equivalent orientations for the global minimum, at $\phi$=180$^{\circ}$ and $\phi$=300$^{\circ}$.
We illustrate in Fig.~\ref{3D-PES} (left panel), the two-dimensional cut of the interaction potential as a function of two Jacobi coordinates ($R$ and $\theta$) while the $\phi$ angle is held fixed at its equilibrium
value corresponding to the CH$_3$CCH-He minimum.
The variation in this cut shows a strong anisotropy of the interaction potential of CH$_3$CCH-He along the $\theta$ coordinate.
For a better appreciation of the topography of the PES, we present also 2D-cut of the 3D-PES as a function of $\theta$ and $\phi$ for $R=6.6$ bohr (Fig.~\ref{3D-PES} right panel). This type of plot offers a unique overview as it includes all minima and the barriers between them.
\section{Scattering calculations}\label{section:3}
CH$_3$CCH is a prolate symmetric top molecule that belongs to the C$_{3v}$ point group with a dipole moment of 0.7829 Debye \citep{wlodarczak1988submillimeter}.
The relevant rotational Hamiltonian associated with CH$_3$CCH is expressed as follow:
\begin{equation}
 H_{rot}=\frac{\hbar^2}{2I_b}j^2+\hbar^2(\frac{1}{2I_a}-\frac{1}{2I_b})j_a^2
\end{equation}
where $j^2$ is the square of the angular momentum that satisfies the relation $j^2$=$j^2_a+j^2_b+j^2_c$ and $I_a$ and $I_b$ are the principal moments of inertia.
The wave functions $\vert jkm \rangle $ of a symmetric top molecule are defined by three quantum numbers, where $k$ denotes 
the projection of $j$ along the $C_{3v}$-axis of the body-fixed reference and $m$ is its projection on the $z$-axis of the space-fixed frame of reference.
The energies of the rotational levels are given by :
\begin{equation}
E_{j,k}=Bj(j+1)+(A-B)k^2
\end{equation}
where $B$ and $A$ are the rotational constants which are equal to 0.2850 cm$^{-1}$ and 5.3083 cm$^{-1}$ \citep{el2001vibrational} respectively.
Due to the identical nuclear spin of hydrogen atoms, we distinguish two independent forms of propyne: $ortho-$CH$_3$CCH (or, $A-$) and $para$-CH$_3$CCH (or, $E-$).
The rotational levels associated to the propyne molecule are characterized by $k= 3n\pm 1$ for $para-$CH$_3$CCH and $k= 3n$ for $ortho-$CH$_3$CCH ($n$ being integer).\\
It is important to note that the \textit{ortho} to \textit{para} transitions are strictly forbidden in such non-reactive collisions. Therefore, the scattering computations can be carried out separately for each nuclear spin species.
The small values of the rotational constants $A$ and $B$ lead to a complex rotational structure with a high density of rotational levels. In the ISM, CH$_3$CCH is usually detected via transitions between highly excited states, e.g. 21$_k$ to 20$_k$, $k$= 0,...,7 \citep{calcutt2019alma}. This implies that cross sections of CH$_3$CCH-He need to be carried out using a large rotational basis set, which limit the possibility of employing a full quantum-mechanical dynamics methods for scattering calculations.

\subsection{Cross sections}
The main focus of this paper is the study of the collisional (de-)excitation of CH$_3$CCH by He atoms.
Cross sections for rotationally inelastic transitions of $para$ and $ortho-$CH$_3$CCH-He were carried out using the time-independent close-coupling (CC) method up to $E_{\text{tot}} \leq$ 100 cm$^{-1}$ and the coupled state (CS) approximation for total energies ranging from 100 $\leq$ $E_{\text{tot}} \leq$ 600 cm$^{-1}$.\\
The scattering calculations were carried out using the \cite{manolopoulos1986improved} propagator to solve the coupled differential equations implemented in the MOLSCAT computer code.\\
To keep computational time under control, we start by testing the integrator parameters, \textit{STEPS} and the size of the rotational basis set $N_{\text{lev}}$, which were adjusted to ensure convergence of the rotational cross-sections over the entire energy range.
Hence, the integration limits of the scattering computations were fixed at $R_{\text{min}}$=3.5a$_0$ and $R_{\text{max}}$=50a$_0$.
The \textit{STEPS}-parameter is set to 100 for $E_{\text{tot}} \leq$ 50 cm$^{-1}$, 70 for $E_{\text{tot}} \in$ [50,100] cm$^{-1}$, 50 for 100 $\leq E_{\text{tot}} \leq$ 200 cm$^{-1}$ and 30 for $E_{\text{tot}} \in$ [200,600] cm$^{-1}$. All these propagator parameters were fixed by carrying out convergence tests.
Furthermore, the size of the rotational basis set was optimized in order to include in the scattering computations all open channels along with a few closed channels.

\begin{figure*}
\centering
{\label{c}\includegraphics[width=.49\linewidth]{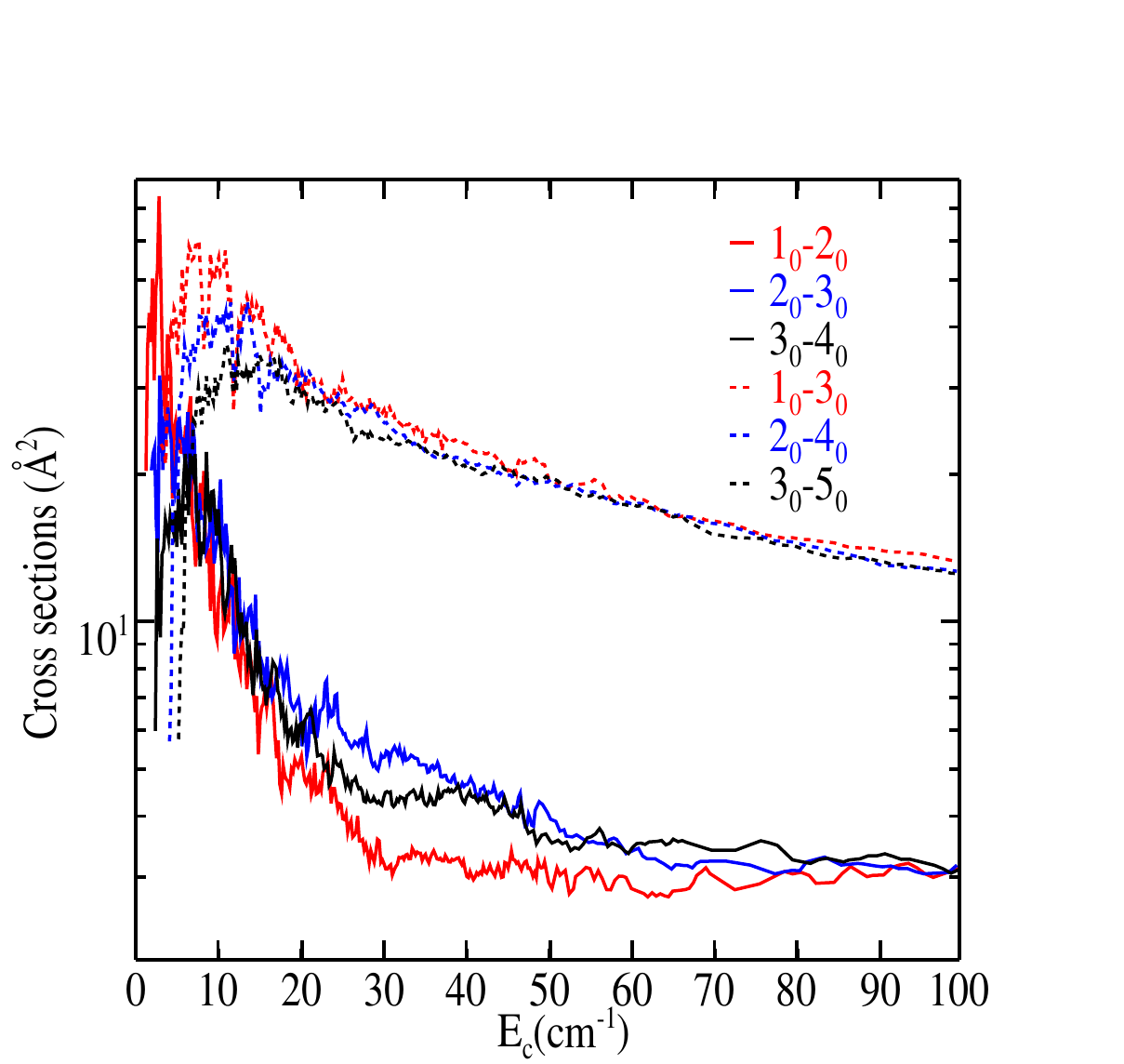}}
{\label{c}\includegraphics[width=.48\linewidth]{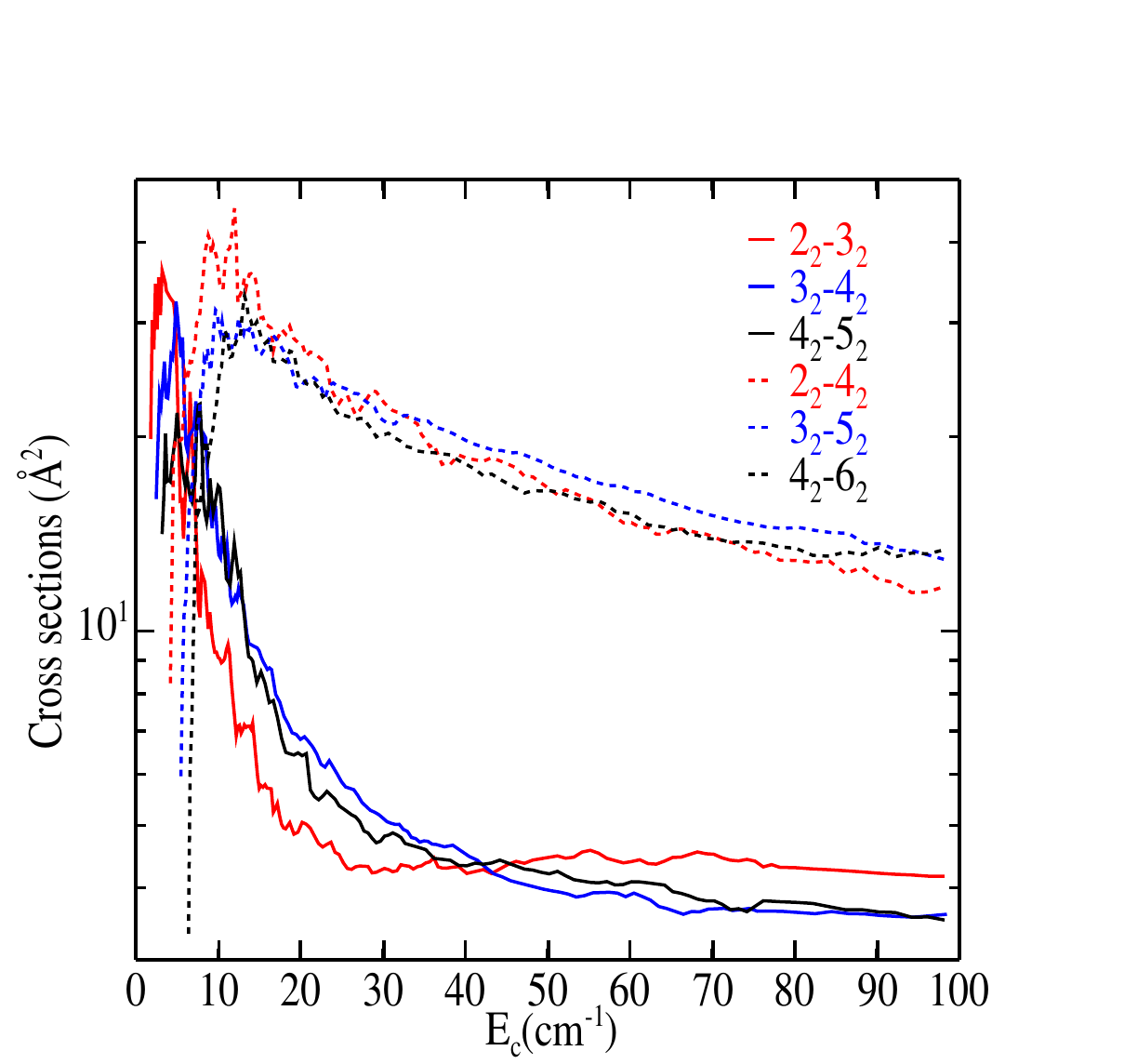}}
\caption{Kinetic energy dependence of the rotational excitation cross sections $j_{k} \rightarrow j'_{k'}$ of \textit{ortho}-CH$_3$CCH-He (left panel) and \textit{para}-CH$_3$CCH-He (right panel) in collision with He for $\Delta j=1$ and $\Delta j=2$ transitions while $\Delta k=0$.}
\label{XCSS}
\end{figure*}

For $ortho-$CH$_3$CCH, the number of rotational levels $N_{\text{lev}}$ considered for our calculations was taken as : $N_{\text{lev}}$=55 (up to $j_k$=24$_3$ with an energy of 216.24 cm$^{-1}$) for total energies $E_{\text{tot}} \leq$ 50 cm$^{-1}$, $N_{\text{lev}}$=80 (up to $j_k$=31$_3$ with an energy of 327.98 cm$^{-1}$) for total energies 50 cm$^{-1} \leq E_{\text{tot}} \leq$ 100 cm$^{-1}$, and $N_{\text{lev}}$=155 (up to $j_k$=49$_0$ with an energy of 698.39 cm$^{-1}$) for 102 cm$^{-1} \leq E_{\text{tot}} \leq$ 600 cm$^{-1}$. For $para-$CH$_3$CCH, we take $N_{\text{lev}}$=105 (up to $j_k$=11$_7$ with an energy of 283.79 cm$^{-1}$) for total energies $E_{\text{tot}} \leq$ 100 cm$^{-1}$, $N_{\text{lev}}$=160 (up to $j_k$=37$_2$ with an energy of 420.88 cm$^{-1}$) for total energies 100 $\leq$ $E_{\text{tot}} \leq$ 200 cm$^{-1}$, $N_{\text{lev}}$=185 (up to $j_k$=38$_4$ with an energy of 502.83 cm$^{-1}$) for total energies 202 $\leq$ $E_{\text{tot}} \leq$ 300 cm$^{-1}$ and $N_{\text{lev}}$=215 (up to $j_k$=19$_{10}$ with an energy of 610.65 cm$^{-1}$) for total energies 305 $\leq$ $E_{\text{tot}} \leq$ 600 cm$^{-1}$.\\
We attribute a value of 0.05~\AA$^2$ to the off-diagonal tolerance 
which determines the maximum value of the total angular momentum ($J$) that needs to be taken into account.
Finally, given that the total angular momentum $J$ is a quantity preserved, the integral cross section was obtaining by summing partial wave contributions.\\
We calculated rotational cross sections for total energy ranging from 0.1 to 600 cm$^{-1}$. To do so, the energy grid was tuned to describe all the resonances in the cross sections and to cover the entire energy range required to compute collisional rates accurately up to 100 K.
We used an energy step $dE$ of 0.2 cm$^{-1}$ for energies below 50 cm$^{-1}$, 0.5 cm$^{-1}$ for energies between 50 cm$^{-1}$ and 100 cm$^{-1}$, 2 cm$^{-1}$ for energies between 100 cm$^{-1}$ and 200 cm$^{-1}$,  5 cm$^{-1}$ for energies between 200 cm$^{-1}$ and 400 cm$^{-1}$, and  10 cm$^{-1}$ for energies between 400 cm$^{-1}$ and 600 cm$^{-1}$.\\
For kinetic energies where resonances become negligible ($E_{\text{c}} 
\geq$ 100 cm$^{-1}$), the calculations using CC method, in the fully converged limit, turned out to be impractical and
would require an inordinate amount  of computing time, thus we explored using the CS approximation to save time calculation. The CS approximation is indeed expected to be reliable at high kinetic energies \citep{phillips1996collisional}. 
 We examined the accuracy of CS computations by direct comparison to CC ones as presented in Table~\ref{table:1} at a total energy of 100 cm$^{-1}$.
As can be seen in Table~\ref{table:1}, the relative error between CC and CS cross sections for selected transitions does not
exceed 10\% for $E_{\text{tot}}$=100 cm$^{-1}$ with a disk occupancy and CPU time for the CC calculations that are several times larger than CS ones. We conclude that for CH$_3$CCH-He collisions, the CS method can be used for total energies larger than 100 cm$^{-1}$.  \\
\begin{table}
\centering
\caption{Comparison between CC and CS cross sections (in ~\AA$^2$) for the excitation of \textit{para} CH$_3$CCH by He for total energies $E_{\textrm{tot}}=100$ cm$^{-1}$ with $N_{\text{lev}}$= 80.}
\label{table:1}
\begin{tabular}{ccccc}
\hline
\hline
 Energy & $j_{k_k} \rightarrow j'_{k'}$ & CC & CS & Error \\
\hline
 $E=100$ cm$^{-1}$ & 1$_1 \rightarrow 2_1$ & 4.0361 &  3.9202 & 2.8\% \\
                   & 3$_1 \rightarrow 4_1$ & 3.0956 &  3.1104 & 0.4\% \\
                   & 5$_1 \rightarrow 6_1$ & 2.9126 &  3.0719 & 0.6\% \\
                   & 7$_1 \rightarrow 9_1$  & 11.9612 & 11.8952 & 0.5\%\\
                   & 3$_2 \rightarrow 5_2$  & 14.4978 & 15.1471 & 4.4\% \\
                   & 4$_1 \rightarrow 5_2$ & 3.6750 &  3.7966 & 3.3\% \\
                & 9$_1 \rightarrow 10_1$ & 2.7101 & 2.8221 & 7.8\% \\
                & 6$_2 \rightarrow 9_2$ & 1.6051 &  1.7571 & 9.4\% \\
\hline
\hline

\end{tabular}
\end{table}
Figure~\ref{XCSS} shows examples of the kinetic energy dependence of collisional cross sections for the rotational excitation of \textit{para-}CH$_3$CCH–He (right panel) and \textit{ortho-}CH$_3$CCH–He (left panel) for some dipolar ($\Delta j$ = 1) and quadrupolar ($\Delta j$ = 2) transitions with $\Delta k$ = 0.
These figures underline the  importance of using a sufficiently fine energy grid to correctly describe the resonances, especially for $E_{\text{c}} \leq$ 50 cm$^{-1}$. The cross sections for these transitions increase above the threshold at which these transitions open until achieving a maximum and then decrease as the collision energies increase. 
The cross-sections exhibit many Feshbach and shape resonances at kinetic energies below 50 cm$^{-1}$. This is due to the formation of quasi-bound states into the potential well depth before dissociation of the collisional complex. \\
Figure~\ref{XCSS} reveals the existence of a strong even $\Delta j$ propensity rule at almost all collision energies for CH$_3$CCH-He. The largest cross sections (in magnitude) are found for transitions with $\Delta j$=2 that is, 1$_0$-3$_0$, 2$_0$-4$_0$ and 3$_0$-5$_0$ for \textit{ortho}-CH$_3$CCH-He and for the transitions 1$_1$-3$_1$, 2$_1$-4$_1$, and 3$_1$-5$_1$ for \textit{para}-CH$_3$CCH-He.
These propensity rules can be explained by exploring the shape of the PES, as the symmetry of the interaction potential will promote transitions with even |$\Delta j$|. Furthermore, this propensity rule can be understood by examining the radial terms $V_{lm}$(R) described in Fig.~\ref{coeff}.
Cross sections with |$\Delta j$| = 2 are mainly caused by the $V_{20}$ term, which is the dominant anisotropic term for CH$_3$CCH-He, hence, the magnitude of cross section corresponding to transitions with |$\Delta j$| = 2 are larger than for other transitions.\\
Propensity rules that favor even values of $\vert \Delta j \vert $ were already observed for the excitation of other symmetric top molecules with C$_{3v}$ symmetry by He atoms,  such as CH$_3$CN\citep{ben2022interaction} and SiH$_3$CN\citep{naouai2021inelastic}.
Figure~\ref{XCSS} only illustrates cross sections for \textit{para} and \textit{ortho}-CH$_3$CCH-He and for transitions with $\Delta j$=0, the same behavior is also observed for other values of $\Delta k$.
\section{Rate coefficients}\label{section:4}
\begin{figure}
\centering
	\includegraphics[width=0.99\columnwidth]{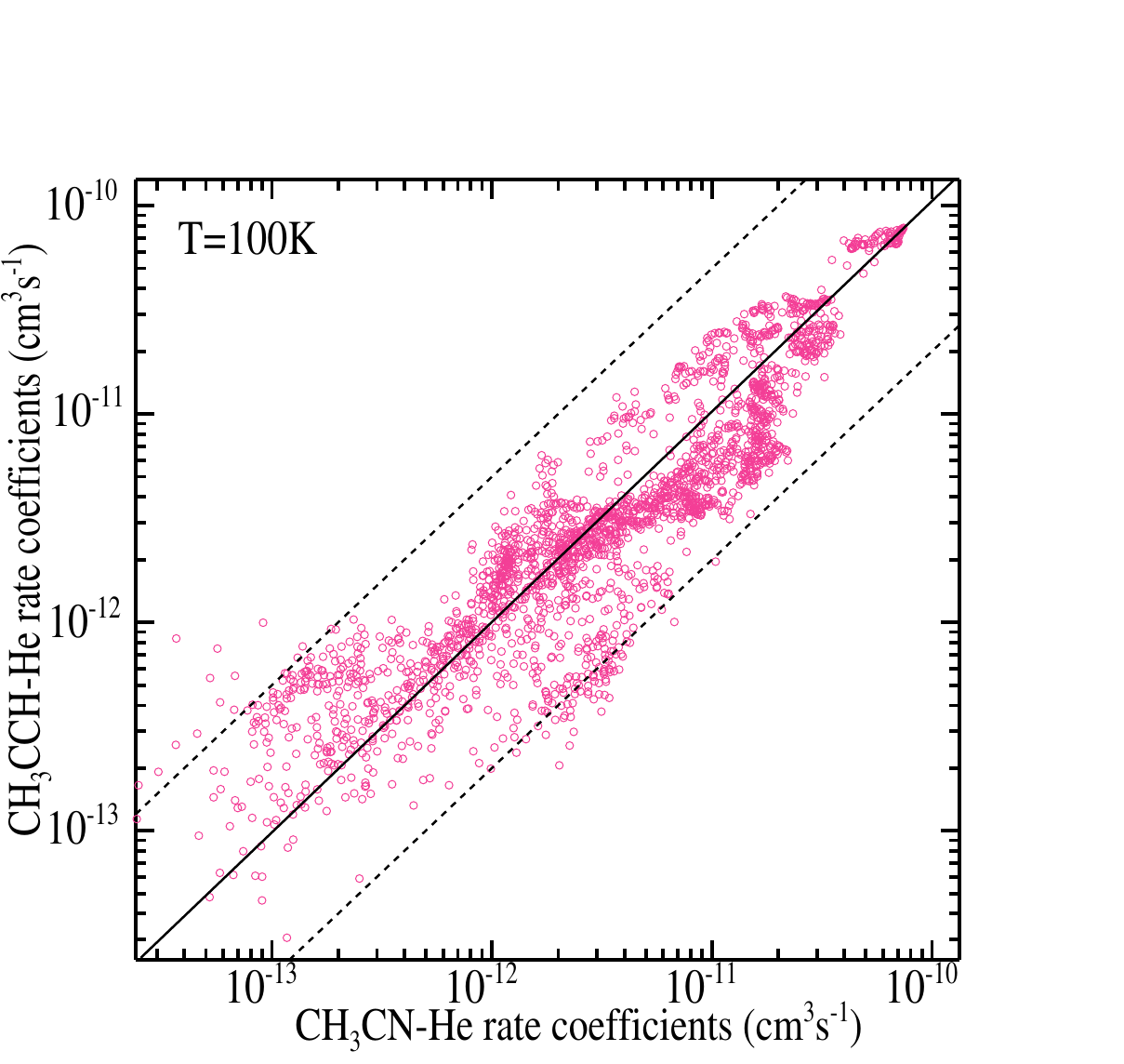}
	\caption{Comparison between \textit{ortho} and \textit{para} CH$_3$CCH-He and CH$_3$CN-He rate coefficients at T = 100 K. The
diagonal line corresponds to equal rate coefficients and the dashed line to the difference of a factor of 5.}\label{CN-CH}
\end{figure}

Calculated state-to-state rotational cross sections were used to compute  the corresponding rate coefficients for the collisions of \textit{ortho}- and \textit{para}- type levels of methyl acetylene with He atoms by averaging over the collision energy $(E_{\text{c}})$.
\begin{equation}
 k_{i \rightarrow f}(T)=\biggl(\frac{8}{\pi\mu\beta}\biggl)^{\frac{1}{2}}\beta^2\int_0^{\infty} E_c \sigma_{i \rightarrow f}(E_c)e^{-\beta E_c} dE_c
\end{equation}
where $\beta$=$1/k_BT$ and $k_B$, $T$ and $\mu$=3.638773549 au are the Boltzmann constant, the kinetic temperature and the collision reduced mass, respectively. \\
In this work, the state-to-state cross sections calculated for total energy up to 600 cm$^{-1}$ can be used to compute rate coefficients for transitions among the first 60 \textit{ortho} levels (up to $j_k$=13$_6$, $E_{\text{rot}}$=232.554 cm$^{-1}$) and 60 \textit{para} levels (up to $j_k$=10$_5$, $E_{\text{rot}}$=156.862 cm$^{-1}$) for kinetic temperature between 5 and 100 K. This complete
set of (de-)excitation rate coefficients will be made available online
on the EMAA (\url{https://emaa.osug.fr} and \url{https://dx.doi.org/10.17178/EMAA}) and BASECOL \citep{dubernet2023basecol} databases for line radiative transfer studies.\\
Figure~\ref{section:5} displays plots of the quenching rate coefficients of CH$_3$CCH-He as a function of kinetic temperature for
selected $\Delta j$ = 2 transitions accompanied by $\Delta k$ = 0, 1 and 3. These plots show that
the order of magnitude of rate coefficients associated to transitions with $\Delta k$ = 0, is larger than those with $\Delta k$ > 0. The largest rate coefficients are found
for the transitions : 5$_3$-3$_3$, 6$_3$-4$_3$ and 7$_3$-5$_3$ for \textit{ortho}-CH$_3$CCH-He and 4$_2$-2$_2$, 5$_2$-3$_2$ and 6$_2$-4$_2$ for \textit{para}-CH$_3$CCH-He. 
The rate coefficients for $\Delta j =2$ transitions are seen to vary only weakly with temperature in the range investigated here. However, for $\Delta j =1$ transitions (not shown), a stronger variation is observed with a decrease of the rate coefficients from 5 K to about 30 K, and a slow increase at higher temperature.
To conclude, transitions with even |$\Delta j$| and |$\Delta k$|=0 are more favorable
compared to those with |$\Delta j$| and |$\Delta k$| $\ne$ 0. These results correspond to those already observed at the level of the cross sections (see Fig.~\ref{XCSS}).\\
To the best of our knowledge, no collisional rates for the excitation of
CH$_3$CCH were available in literature and the rate coefficients for CH$_3$CN were sometimes used in radiative
transfer models to approximate the excitation of propyne in astrophysical media \citep{mauersberger1991dense}. To assess the validity of such an approach in the collisional studies, we performed a comparison of CH$_3$CCH-He and CH$_3$CN-He collisional rate coefficients, computed recently by \cite{ben2023collisional} with the same level of theory for both the construction of the PES as well as for the scattering dynamics. We present this comparison in Fig.~\ref{CN-CH} for transitions between the first 50 \textit{ortho} levels and 50 \textit{para} levels of CH$_3$CCH and CH$_3$CN at temperature $T$ = 100 K. One can see that collisional rate coefficients for the dominant transitions ($k$ > 10$^{-11}$ cm$^3$s$^{-1}$, corresponding to |$\Delta k$| = 0 transitions) of both species agree within a factor of 2, however, we observe important differences for the smaller collisional rates with relative differences that reach a factor of 5. These differences show that independent calculations have to be performed for each molecule as the
rate coefficients cannot be assumed to be equal.\\
The collisional rate coefficients obtained for the CH$_3$CCH-He and CH$_3$CN-He species allow us to predict that their excitation
in the astrophysical environments is different. Therefore, a radiative transfer study
must be carried out for CH$_3$CCH using its respective set of collisional rate coefficients to obtain an accurate abundance of propyne.\\
It is important to emphasize that this comparison is only for collisional excitation by He atoms, while the dominant collider in the ISM is H$_2$ molecule. At present, no collisional data is available for either CH$_3$CCH-H$_2$ or CH$_3$CN-H$_2$. In this case, one usually scales all rate coefficients for excitation by He atoms by a single factor to account for the mass difference between He and H$_2$, but the accuracy of such an approach is known to be limited.

\begin{figure*}
\centering
{\label{c}\includegraphics[width=.49\linewidth]{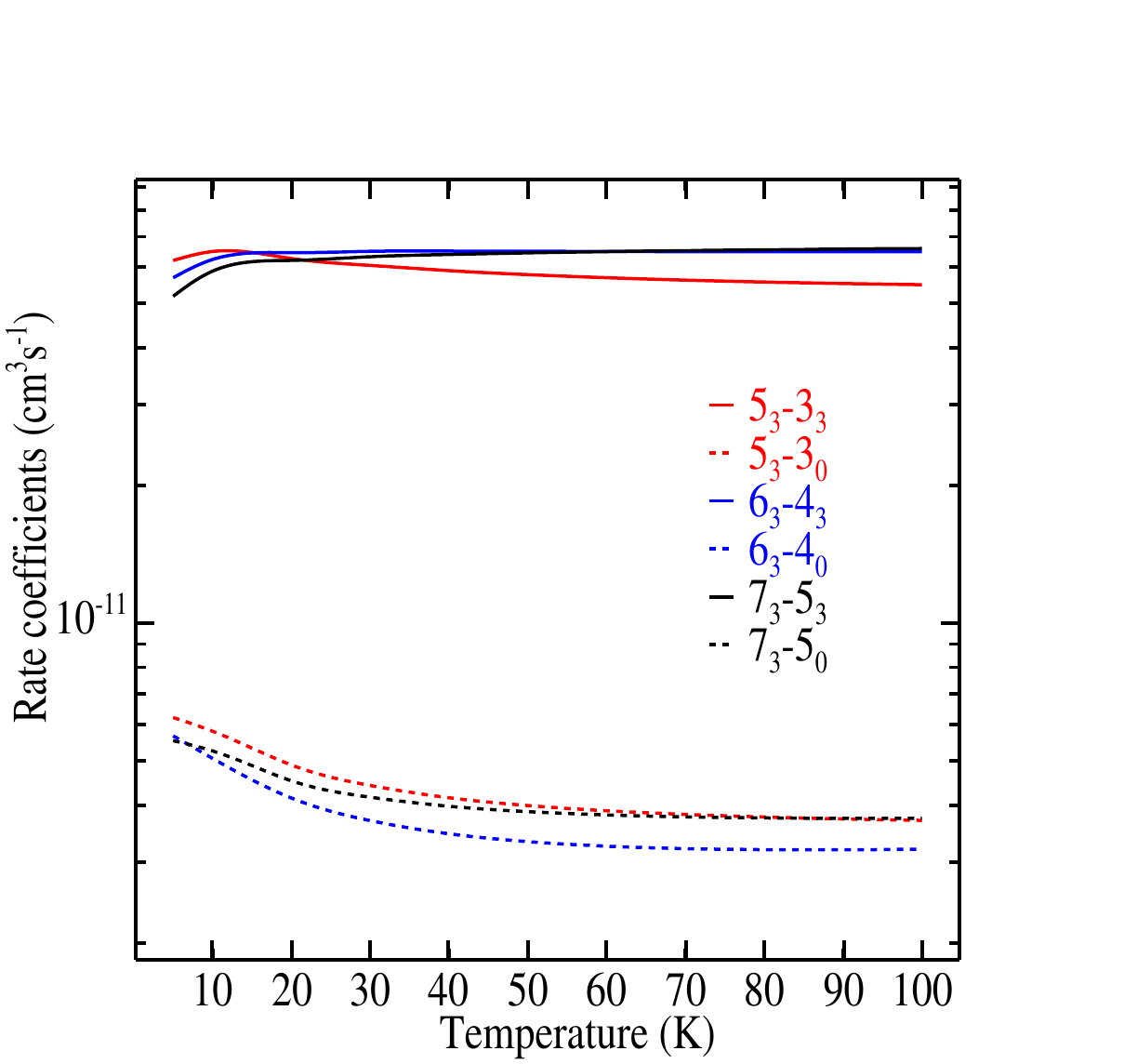}}
{\label{c}\includegraphics[width=.49\linewidth]{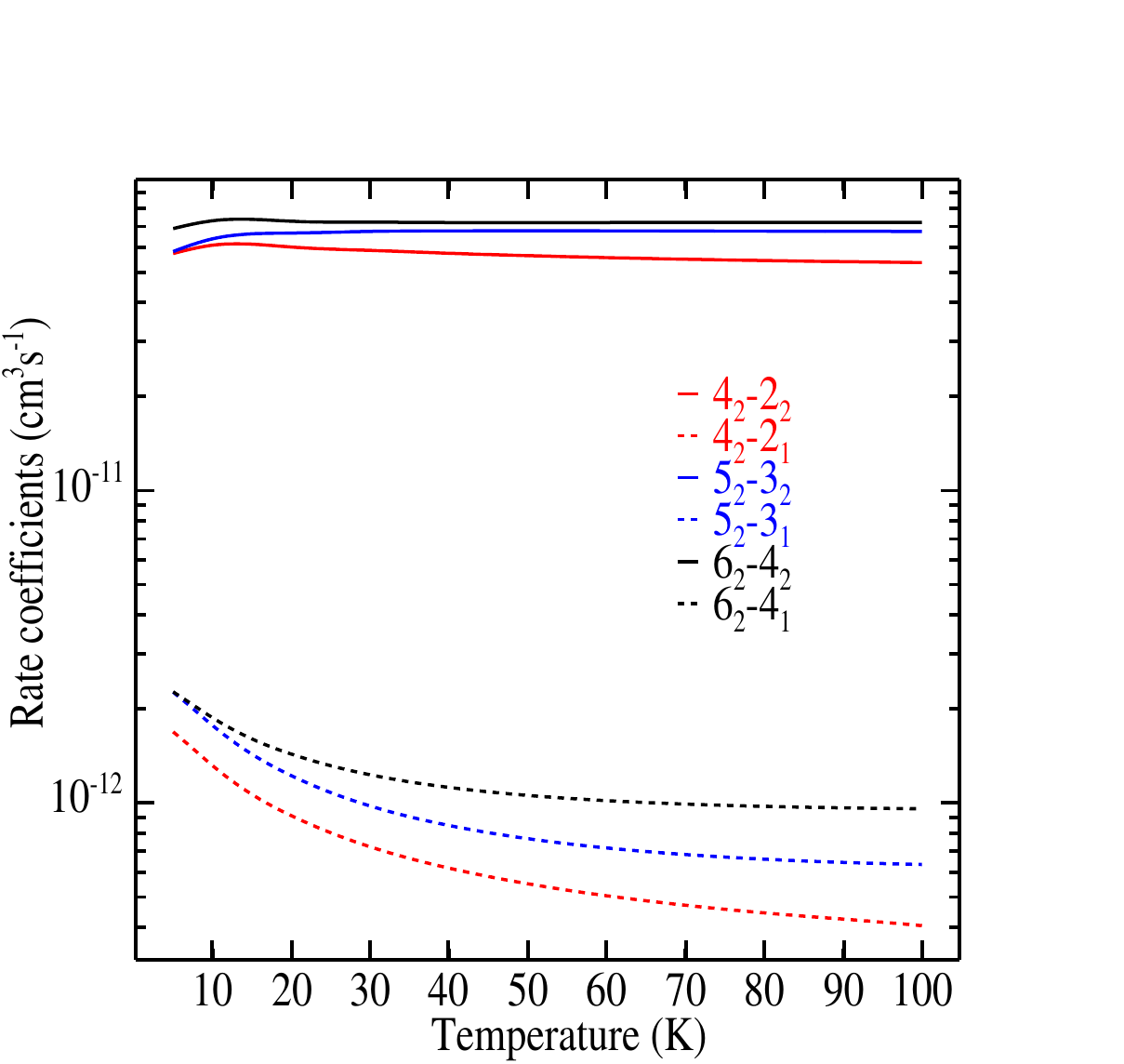}}
\caption{Temperature dependence of the rotational excitation rate coefficients $j_{k} \rightarrow j'_{k'}$ of \textit{ortho}-CH$_3$CCH-He (left panel) and \textit{para}-CH$_3$CCH-He (right panel) in collision with He atom for $\Delta j=2$ and $\Delta k=0, 1$ and 3.}
\label{XCS}
\end{figure*}

\section{Application}\label{section:5}
\begin{figure*}
\centering
{\label{a}\includegraphics[width=.33\linewidth]{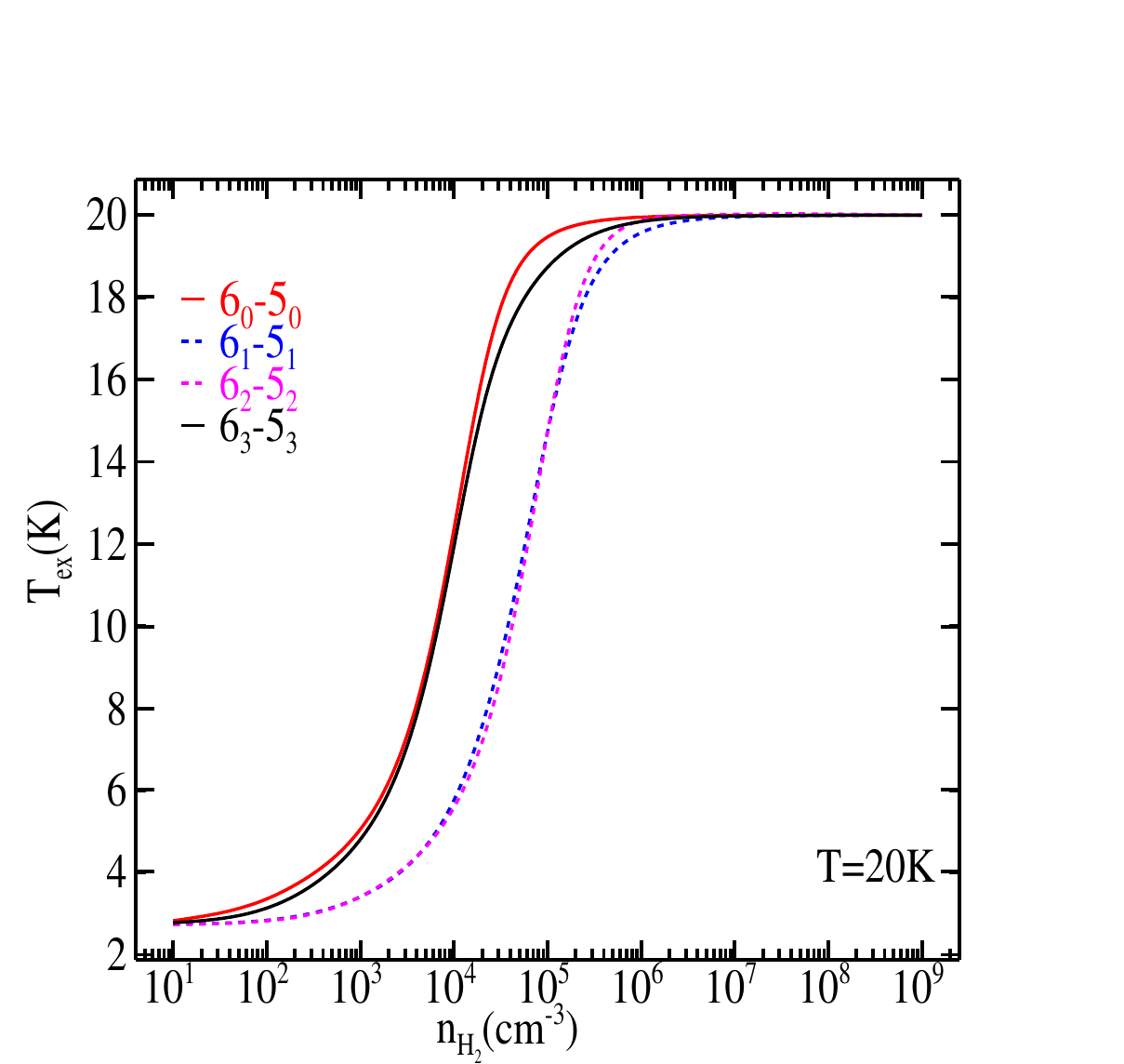}}
{\label{b}\includegraphics[width=.33\linewidth]{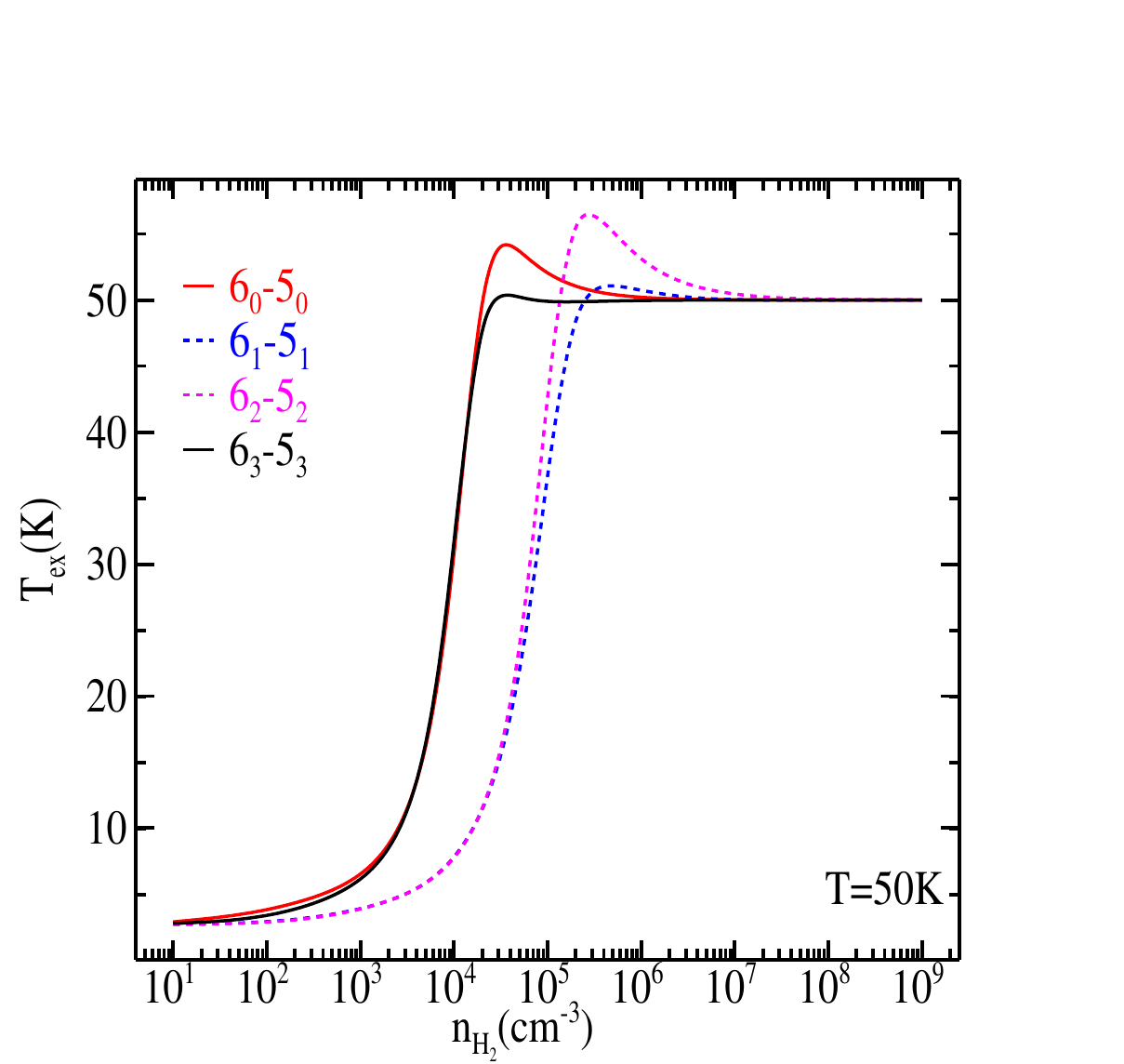}}
{\label{c}\includegraphics[width=.33\linewidth]{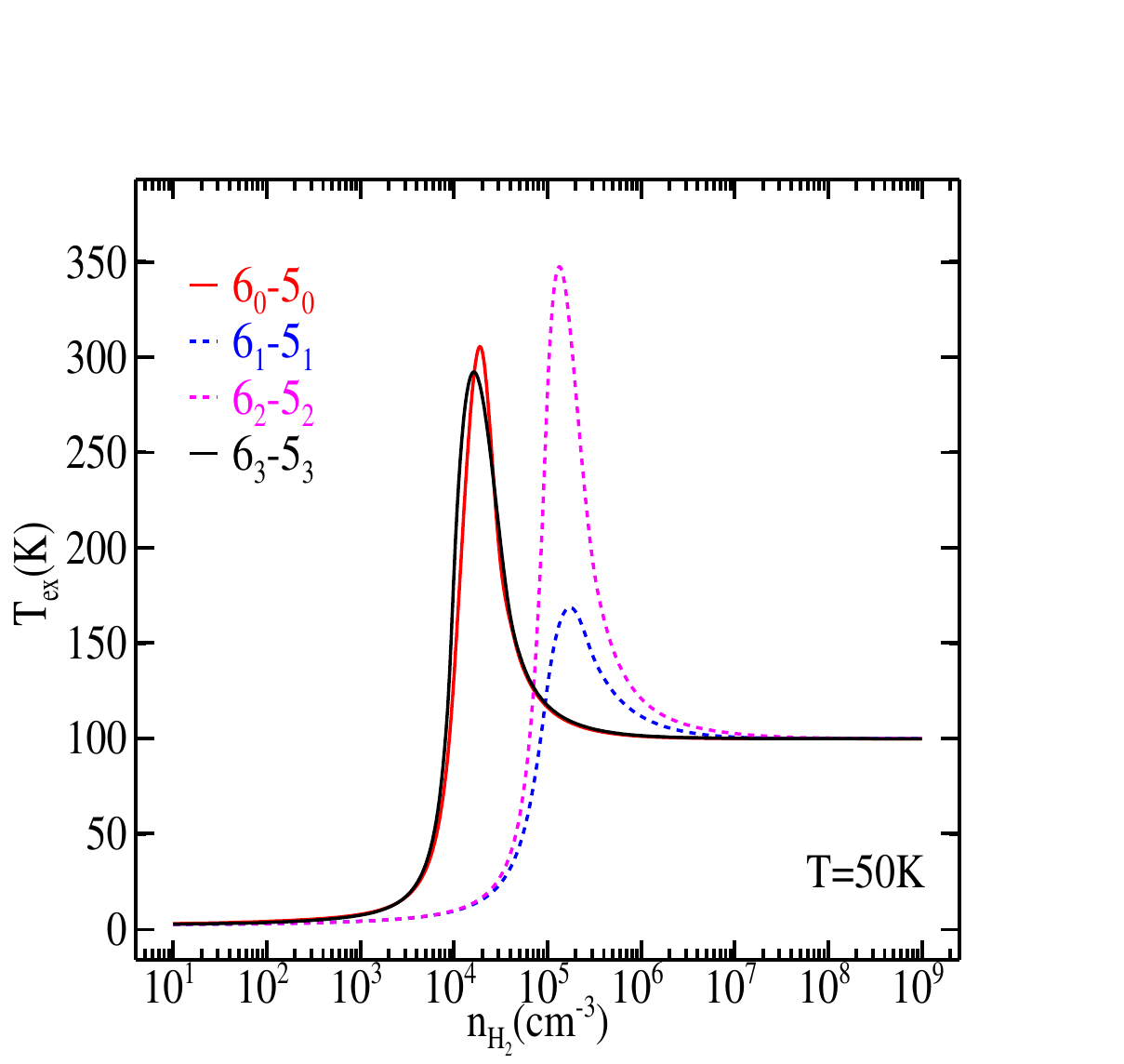}}
\caption{Excitation temperature of CH$_3$CCH for the transitions
$j_k \rightarrow j^\prime_{k^\prime}$ with $j=6$, $j^\prime=5$ and $k=k^\prime=0$ or 2 ($para$-CH$_3$CCH) and $k=k^\prime=0$ or 3 ($ortho$-CH$_3$CCH)
as a function of the H$_2$ density for three kinetic temperature (20 K, 50 K and 100 K) and a column density of 10$^{14}$ cm$^{-2}$. }
\label{application-ex} 
\end{figure*}

In the interstellar medium, the intensities of the emission lines are proportional to the population of the energy levels of the molecule. The evolution of these populations is based on collisional and radiative processes that are governed by collisional rate coefficients and radiative Einstein coefficients, respectively. The knowledge of these two sets of data allows one to perform a radiative transfer computation for the propyne molecule using the Radex \citep{van2007computer} computer code.\\
Non-LTE radiative transfer computations were performed using the escape probability approximation. Our collisional rate coefficients (from 5 to 100
K) were combined with spectroscopic data from the Cologne Data base for Molecular Spectroscopy portal \citep{muller2005cologne}. In practice, we have computed the excitation ($T_{\text{ex}}$) and brightness ($T_{\text{B}}$) temperatures for the transitions $j_k$=6$\rightarrow$4 with $k$=0,1,2 and 3 at 102.5 GHz which are usually detected in astrophysical clouds \citep{andron2018methyl}. To do so, we scaled the CH$_3$CCH-He rates by a factor of 1.39 to approximate the rate coefficients that \textit{para-}H$_2$($j$=0) would induce and  we fixed the cosmic microwave background, the line width and the column density of propyne at 2.73 K, 2.1 km.s$^{-1}$ \citep{andron2018methyl} and 10$^{-14}$ cm$^{-2}$, respectively. Finally, we vary the volume density of the molecular hydrogen from 10$^2$ cm$^{-3}$ to 10$^8$ cm$^{-3}$ to examine the behaviour of the excitation of CH$_3$CCH under and out of LTE conditions.\\
We display in Fig.~\ref{application-ex}, the variation of 
the excitation temperature of CH$_3$CCH for selected kinetic 
temperatures of 20, 50 and 100 K. We observe a supra-thermal 
effect at $T=$ 50 and 100 K for H$_2$ densities between 
10$^4$ and 10$^6$ cm$^{-3}$, where the excitation temperature 
is larger than the kinetic temperature. At low H$_2$ density 
($n_{H_2}$ < 10$^4$ cm$^{-3}$), the medium is dilute and the excitation 
temperature is given by the cosmic background temperature 
($T_{\text{ex}}$ = $T_{\text{CMB}}$ = 2.73 K), while at high 
density ($n_{H_2}$ > 10$^6$ cm$^{-3}$), when the collisional excitation processes 
becomes more important, the 
$T_{\text{ex}}$ tends asymptotically to the kinetic 
temperature where LTE conditions are reached. From these 
plots, we observe that the LTE is reached for volume 
densities above 10$^6$ cm$^{-3}$.\\

\section{Conclusion}\label{section:6}
In this paper, the first 3D-PES for the interaction between CH$_3$CCH and He atoms was computed using the explicitly correlated coupled cluster theory [CCSD(T)-F12a] and the aug-cc-pVTZ basis set. The global minimum was found to correspond to a potential well depth of 51.04 cm$^{-1}$ and an equilibrium intermolecular separation of 6.3 bohr.
This interaction potential was used to compute rotationally-inelastic cross sections for transitions between the first 60-\textit{ortho} and 60-\textit{para} levels of CH$_3$CCH in collision with He atoms in time-independent close-coupled and coupled states quantum scattering computations. 
Rotational cross sections were calculated as a function of the total energy up to 600 cm$^{-1}$.\\
By thermally averaging the cross sections over the collision energy,
collisional rates were computed for temperatures relevant to interstellar conditions, from 5 K to  100 K.\\
We have compared the new set of collisional rates of CH$_3$CCH-He with those for CH$_3$CN-He. This comparison shows that collisional rate coefficients are different by up to a
factor of 5. Consequently, we must use the appropriate collisional rates
of CH$_3$CCH to study the excitation of propyne in the interstellar medium.
We therefore recommend the use of these set of collisional rates in non-LTE models of CH$_3$CCH excitation.\\
We have also used the new set of rate coefficients in a simple radiative transfer model to assess their impact on four observed transitions. This study shows that the propyne lines are not thermalized and 
non-LTE models should be employed to analyse its emission spectra.

\begin{acknowledgements}
      MBK thanks H. da Silva, Jr for help with HPC computing. JL. acknowledges support from KU Leuven through grant no. C14/22/082.
The scattering calculations presented in this work
were performed on the VSC clusters (Flemish Supercomputer Center), funded by the Research Foundation-Flanders (FWO) and the Flemish Government.
\end{acknowledgements}

%
%

\end{document}